\documentclass[14pt,preprint2,longabstract]{aastex}
\newcounter{column_number}
\shorttitle{A {\it Chandra} survey of Galactic globular clusters}
\shortauthors{Cheng et al.}
\begin{document}
\title{Exploring the Mass Segregation Effect of X-ray Sources in Globular Clusters: The Case of 47 Tucanae}
\author{Zhongqun Cheng$^{1,2,3}$, Zhiyuan Li$^{2,3}$, Xiangdong Li$^{2,3}$, Xiaojie Xu$^{2,3}$ and Taotao Fang$^{1}$}
\affil{$^{1}$ Department of Astronomy and Institute for Theoretical Physics and Astrophysics, Xiamen University, Xiamen, Fujian 361005, China} 
\affil{$^{2}$ School of Astronomy and Space Science, Nanjing University, Nanjing 210023, China} 
\affil{$^{3}$ Key Laboratory of Modern Astronomy and Astrophysics (Nanjing University), Ministry of Education, Nanjing 210023, China}

\email{zqcheng@xmu.edu.cn; fangt@xmu.edu.cn}

\begin{abstract}
Using archival {\it Chandra} observations with a total exposure of 510 ks, we present an updated catalog of point sources for Globular Cluster 47 Tucanae. Our study covers an area of $\sim 176.7$ arcmin$^{2}$ (i.e., with $R\lesssim7.5\arcmin$) with 537 X-ray sources.
We show that the surface density distribution of X-ray sources in 47 Tuc is highly peaked in cluster center, rapidly decreases at intermediate radii, and finally rises again at larger radii, with two distribution dips at $R\sim 100\arcsec$ and $R\sim 170\arcsec$ for the faint ($L_{X}\lesssim 5.0\times 10^{30} {\rm\ erg\,s^{-1}}$) and bright ($L_{X}\gtrsim 5.0\times 10^{30} {\rm\ erg\,s^{-1}}$) groups of X-ray sources, separately. 
These distribution features are similar to those of Blue Straggler Stars (BSS), where the distribution dip is located at $R\sim 200\arcsec$ \citep{ferraro2004}. 
By fitting the radial distribution of each group of sources with a ``generalized King model", we estimated an average mass of $1.51\pm0.17\ M_{\odot}$, $1.44\pm0.15\ M_{\odot}$ and $1.16\pm0.06\ M_{\odot}$ for the BSS, bright and faint X-ray sources, respectively.
These results are consistent with the mass segregation effect of heavy objects in GCs, where more massive objects drop to the cluster center faster and their distribution dip propagates outward further. 
Besides, the peculiar distribution profiles of X-ray sources and BSS are also consistent with the mass segregation model of binaries in GCs, which suggests that in addition to the dynamical formation channel, primordial binaries are also a significant contributor to the X-ray source population in GCs.
\end{abstract}

\section{Introduction}
Globular Clusters (GCs) are ancient stellar systems that evolve with some fundamental dynamical processes taking place on timescales shorter than (or comparable to) their absolute age, which make them a unique laboratory for learning about two-body relaxation, mass segregation, stellar collisions and mergers, and gravitational core collapse \citep{heggie2003}. 
Among all of these dynamical interactions, the two-body relaxation plays a fundamental role in driving cluster evolution, as it dominates the transportation of energy and mass in GCs. Stars are driven to reach a state of energy equipartition by two-body relaxation, massive stars (or binaries) therefore tend to lose energy and drop to the lower potential well of GCs. Whereas for lower mass stars, they tend to obtain energy and move faster, and will migrate outwards and even escape from the host clusters \citep{heggie2003}.  
The concept of core collapse, linked to the instability caused by the negative heat capacity of all self-gravitating systems, was first investigated theoretically in the 1960s and confirmed observationally in the 1980s (see \citet{meylan1997} for a review).
During the phase of core-collapse, the stellar density in the cluster core may increase by several orders of magnitude, which significantly increases the frequency of interactions and collisions between stars.

Binaries are thought to play an essential role in the evolution of GCs, as they have much larger cross section and hence much higher encounter frequency than the single stars in GCs \citep{hut1992}. More importantly, binaries in GCs may serve as the reservoir of energy: encounters of binaries in GCs will obey the Hills-Heggie law, soft binaries (with bound energy $|E_{b}|$ less than the averaged kinetic energy $E_{k}$ of the GC stars) tend to absorb energy and become softer or be disrupted, while hard binaries (with $|E_{b}|> E_{k}$) tend to transfer energy to passing stars and become harder \citep{hills1975,heggie1975,hut1993}. Encounters of hard binaries in GCs can thus strongly influence the cluster evolution---sufficient to delay, halt, and even reverse core collapse \citep{heggie2003}. 

Observationally, many exotic objects have been detected in GCs, including Blue Straggler Stars (BSS) \citep{sandage1953}, low-mass X-ray binaries (LMXBs) \citep{clark1975,katz1975}, millisecond pulsars (MSPs) \citep{camilo2005,ransom2008}, cataclysmic variables (CVs) and coronally active binaries (ABs) \citep{grindlay2001,pooley2002a,edmonds2003a,edmonds2003b,heinke2005}. All of these objects either are experiencing the drastic binary evolution stage (i.e., LMXBs, CVs, ABs) or are the immediate remnants of close binaries (i.e., BSS, MSPs). 
Compared with normal binaries, exotic objects hosted in GCs are brighter in luminosity (either in optical, X-ray or in radio band) and can be easily detected and picked out from the dense core of clusters, making them ideal tracing particles of studying stellar dynamical interactions and cluster evolution.  

For example, in searching of BSS in GCs, \citet{ferraro1993} found that the radial distribution of BSS in M3 are bimodal, with two populations of BSS located at cluster center and larger radii, separately, and a region devoid of BSS exists at the medium radii. 
Furthermore, \citet{ferraro2012} found that GCs can be classified into three families (i.e., Family I, II and III) with increasing dynamical ages. 
In this scenario, BSS is flatly distributed with respect to the radial distribution of normal stars in dynamically young (Family I) GCs. However, two-body relaxation will drives the BSS sedimentation toward the cluster center, modifying the flat BSS distribution into a bimodal shape, with a central peak, a dip and an outer rising branch in intermediate dynamical age (Family II) GCs. As time goes on, the radial distribution dip will propagate outward gradually, and eventually leading to a unimodal BSS distribution that monotonically moves outward in dynamically old (Family III) GCs. Based on these findings, \citet{ferraro2012} suggested that BSS in GCs can be utilized to build a ``dynamical clock" of evaluating cluster evolution.

For X-ray sources detected in GCs, majority of them are found to be CVs and ABs, while few have been identified as LMXBs and MSPs (see \citet{heinke2010} for a review). 
These objects are suggested to be dynamically formed in GCs \citep{pooley2003,pooley2006}, and the abundances (i.e., number per unit stellar mass) of LMXBs and MSPs are found to be orders of magnitude higher in GCs than in the Galactic field \citep{clark1975,katz1975,camilo2005,ransom2008}. Meanwhile, the cumulative X-ray emissivity (i.e., X-ray luminosity per unit stellar mass) of many GCs are found to be slightly lower than that of the solar neighborhood stars and the dwarf elliptical galaxies \citep{ge2015,cheng2018a}. This suggests a dearth rather than over-abundance of CVs and ABs in most GCs relative to the field.
To explain this contrast, \citet{cheng2018a} argued that stellar interactions (i.e., binary-single or binary-binary encounters) in GCs are efficient in destroying binaries, and a large fraction of soft primordial binaries have been disrupted in GCs before they can otherwise evolve into CVs and ABs. 
For the remaining hard primordial binaries, they are likely being transformed into X-ray-emitting close binaries by stellar interactions, leading to a strong correlation between binary hardness ratio (i.e., the abundance ratio of X-ray-emitting close binaries to main sequence binaries) and stellar velocity dispersion in GCs \citep{cheng2018b}.

However, most of the previous work (i.e., \citealp{pooley2003,pooley2006,cheng2018a,cheng2018b}) on GC X-ray sources were confined within the cluster half-light radius, where stellar density is high and stellar dynamical interactions cannot be ignored.
For X-ray sources located in the cluster halo, they are more likely to descend from primordial binaries and have not experienced strong encounters; however, two-body relaxation is efficient in driving these systems sedimentation to cluster center. This process was ignored in previous work. 
Hence, it remains an open question whether radial distribution of X-ray sources shall be similar to BSS in GCs. 

Globular cluster 47 Tuc is massive ($M\sim 10^{6} M_{\odot}$, \citealp{harris1996}) and of relatively high stellar concentration ($c=2.01\pm0.12$, \citealp{mcLaughlin2006}), making it one of the clusters with the highest predicted stellar encounter rate in the Milky Way \citep{bahramian2013}. Observationally, main sequence binary fraction in 47 Tuc was found to be lower ($f_{b}=1.8\pm0.6\%$) than that of other GCs (with a typical value of $f_{b}\sim1-20\%$, \citealp{milone2012}), which suggests a substantial fraction of primordial binaries had been disrupted or altered by close encounters in this cluster.   
47 Tuc is a typical Family II cluster with a bimodal radial distribution of BSS \citep{ferraro2004}: the BSS distribution dip is located at $170\arcsec<R<230\arcsec$, which is slightly larger than the half-light radius ($r_{h}=3.17\arcmin$, \citealp{harris1996}). 
In X-ray, 47 Tuc is well known for its large number of X-ray sources, many of which have been identified as CVs and ABs \citep{edmonds2003a,edmonds2003b,heinke2005}, quiescent LMXBs \citep{grindlay2001,edmonds2002} and MSPs \citep{bogdanov2006}.
With a deep {\it Chandra} exposure, \citet{bhattacharya2017} presented the largest X-ray sources catalogue for 47 Tuc with 370 X-ray sources. However, due to a smaller X-ray sources searching region (i.e., within $R=2.79\arcmin$), they did not find a significant mass segregation effect of X-ray sources. 

In this work, we present the most sensitive and full-scale {\it Chandra} X-ray point source catalog for 47 Tuc, which covers an area of $\sim 176.7$ arcmin$^{2}$ (i.e., with $R\lesssim 7.5\arcmin$ in 47 Tuc) and has a maximum effective exposure of $\sim510$ ks (same as \citealp{bhattacharya2017}) at the center of the cluster. 
Therefore, the X-ray source catalog presented in our work is much larger than that of \citet{bhattacharya2017} (see Section 3 for details). 
This paper is organized as follows. In Section 2 we present the {\it Chandra} observations and data preparation. 
In Section 3, we describe the creation of the X-ray source catalog. 
We analyze the X-ray source radial distributions in Section 4, and explore its relation to GC mass segregation effect in Section 5. 
A brief summary of our main conclusions is given in Section 6.
  
\section{{\it Chandra} Observations and Data Preparation}
\subsection{Observations and Data Reduction}
The central region of 47 Tuc has been observed 19 times by the {\it Chandra} Advanced CCD Imaging Spectrometer (ACIS) from March 2000 to February 2015. Five observations in March of 2000 were performed with the ACIS-I CCD array at the telescope focus, while the rest were taken with the aim-point at the S3 chip. As listed in Table-\ref{tab:obslog}, 13 observations were performed with a subarry mode, to minimize the effect of pileup for bright sources. 

Starting from the level-1 events file, we used the {\it Chandra} Interactive Analysis Observations (CIAO, version 4.8) tools and the {\it Chandra} Calibration Database (CALDB, version 4.73) to reprocess the data. Following the standard procedures\footnote{http://cxc.harvard.edu/ciao}, we used the CIAO tool {\it acis\_build\_badpix} to create a bad pixel file and employed {\it acis\_find\_afterglow} to identify and flag the cosmic-ray afterglows events. We updated the charge transfer inefficiency, time-dependent gain and pulse height with {\it acis\_process\_events}. Most of the {\it acis\_process\_events} parameters are set as the default values and the VFAINT option is set for certain observations as appropriate. We then removed bad columns, bad pixels and filtered the events file using the standard (ASCA) grades (0,2,3,4,6). We also inspected the background light curve of each observation and removed the background flares using {\it lc\_clean} routine. We further adopted the clean data from chips I0-I3 (for ACIS-I) and chips S2 and S3 (for ACIS-S) for this work. 
 
\subsection{Image Alignment and Merging} 

\begin{figure*}[htp]
\centering
\includegraphics[angle=0,origin=br,height=0.66\textheight, width=1.0\textwidth]{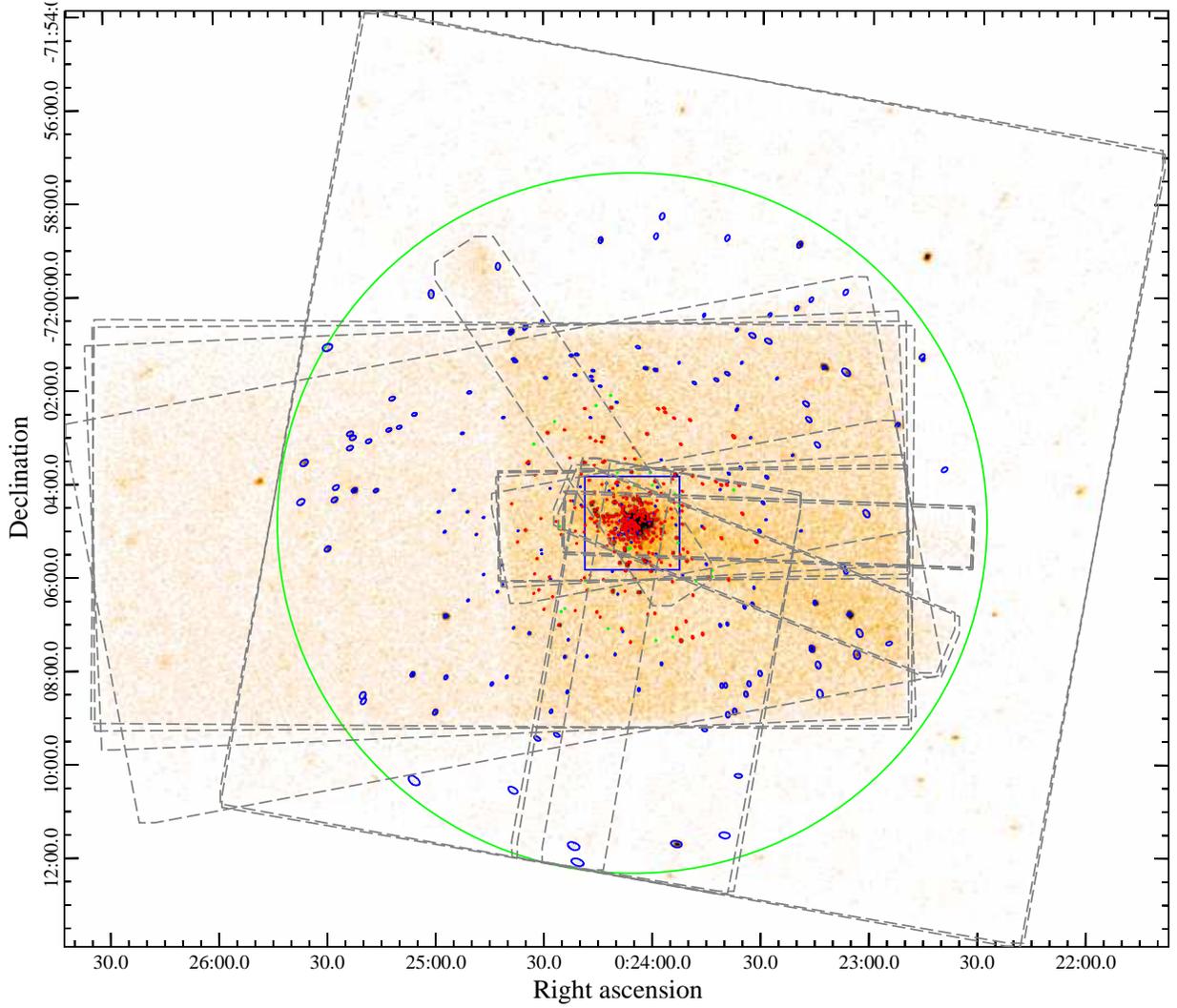}
\linespread{-0.7}
\caption{
{\it Chandra} merged 0.5-7 keV image of 47 Tuc, smoothed with a gaussian kernel with a radius of 5 pixels. Fields of View of the 19 observations in Table-1 are shown as grey dashed boxes. Only X-ray sources located within the green circle ($R=7.5\arcmin$) have been detected and analyzed in this work. The sources shown in red and green were identified by \citet{bhattacharya2017}, while those in blue are new detections in this work. We re-identified those sources in red but failed to identify those in green with our source pruning criterion (see also Section 3.1 for details). A detailed view of the central $2\arcmin \times 2\arcmin$ region (blue square) is illustrated in Figure-\ref{fig:rgbimage}.
 \label{fig:rawimage}}
\end{figure*}

After generating a cleaned event file for each of the 19 observations, we needed to create a merged image for 47 Tuc in order to detect the faintest sources. We created a flux image with a binsize of 0.5 pixel in $0.5-7$ keV band for each observation, and created a PSF map with CIAO tool {\it mkpsfmap} by assuming a power-law spectrum with a photon index of $\Gamma=1.4$ and an enclosed counts fraction (ECF) of 0.4. Supplied with the flux image and PSF map, we then searched X-ray sources within 4 arcmin off-axis of each observation, by using the CIAO tool {\it wavdetect} with a ``$\sqrt{2}$ sequence" wavelet scales (i.e., 1, 1.414, 2, 2.828, 4, 5.656, and 8 pixels) and a false-positive probability threshold of $10^{-6}$. Depending on the exposure time of the observations, $\sim 10-250$ X-ray sources are detected in the individual data sets. In order to improve the alignment between each observation, we refined the position of the detected X-ray sources with the ACIS Extract (AE, \citealp{broos2010}) package, and adopted the centroid positions determined by AE from its ``CHECK\_POSITIONS" stage as the improved positions of these sources. 
AE also provides a tool (i.e., {\it ae\_interObsID\_astrometry}) to verify the astrometric alignment between each observations, which is performed during the pruning stage of the candidate source list (i.e., Section 3.1), to guarantee that the astrometric offsets between two observations are better than $\sim 0.1\arcsec$ \citep{broos2010}.

To correct the relative astrometry among the X-ray observations, we chose the observation (ID:2738) with the longest exposure time as reference of coordinate, and employed the CIAO tool {\it wcs\_match} to match and align the X-ray source lists. We set the search radius of source-pairs to $1\arcsec$ and allow the incrementally elimination of source-pairs with the highest positional errors by setting {\it residtype=1} and {\it residlim=0.5}. The {\it wcs\_match} creates a transformation matrix for each individual observation, with the values of matrix elements range from $-1.168\arcsec$ to $1.195\arcsec$ in linear translation, $-0.084^{\circ}$ to $0.216^{\circ}$ in rotation, and 0.9985 to 1.0019 in scaling (Table-\ref{tab:obslog}). This matrix was used with {\it wcs\_update} to update the aspect solution file, which was then supplied to {\it acis\_process\_events} tool to reprocess the event file. The parameters of {\it acis\_process\_events} are set as described in Section 2.1, except that sub-pixel event repositioning algorithm EDSER\footnote{http://cxc.harvard.edu/ciao/why/acissubpix.html} \citep{li2004} was used to preserve the spatial resolution of point sources on-axis. This algorithm is helpful in improving the resolution of nearby sources in the core of 47 Tuc, where stellar density is high and X-ray sources are crowded. 

We combined these 19 cleaned event files into merged event files and exposure-corrected images with tool {\it merge\_obs}. Three groups of images have been created in soft (0.5--2 keV), hard (2--7 keV) and full (0.5--7 keV) bands and with binsize of 0.5, 1 and 2 pixels, respectively. By setting an ECF of 0.4 and a photon energy of 1.2 keV, 3.8 keV and 2.3 keV for soft, hard and full band, respectively, we also calculated for each of the 19 observations PSF maps with {\it mkpsfmap}, and combined them into exposure time weighted averaged PSF maps with tool {\it dmimgcalc}. These merged images are only used for detecting X-ray sources in this work.

\section{Catalog of Sources}
\subsection{Source Detection} 
To generate the candidate X-ray source list for 47 Tuc, we ran {\it wavdetect} on each of the 9 merged images, using a ``$\sqrt{2}$ sequence" of wavelet scales (i.e., 1, 1.414, 2, 2.828, 4, 5.656, 8, 11.314, and 16 pixels). We confined the X-ray sources searching region as a circle with radius of 7.5 acrmin (Figure-\ref{fig:rawimage}), and adopted a false-positive probability threshold of $1 \times 10^{-7}$, $5 \times 10^{-6}$ and $1 \times 10^{-5}$ for event files with scale of 0.5, 1.0 and 2.0 pixels, respectively. We then combined the {\it wavdetect} results into a master source list with the {\it match\_xy} tool from the Tools for ACIS Review and Analysis (TARA) packages\footnote{http://www2.astro.psu.edu/xray/docs/TARA/}. The resulting source list includes 559 sources. Although the relatively loose source-detection threshold introduces a non-negligible number of spurious sources to the candidate list, we found that dozens of faint sources are failed to be identified by {\it wavdetect} within the crowded core of 47 Tuc, and several sources within bright neighbours also cannot be properly resolved. We picked out these sources by eye and added them into the master source list, which resulting a final candidate list of 617 sources.

\begin{figure*}[htp]
\centering
\includegraphics[angle=0,origin=br,height=0.75\textheight, width=1.0\textwidth]{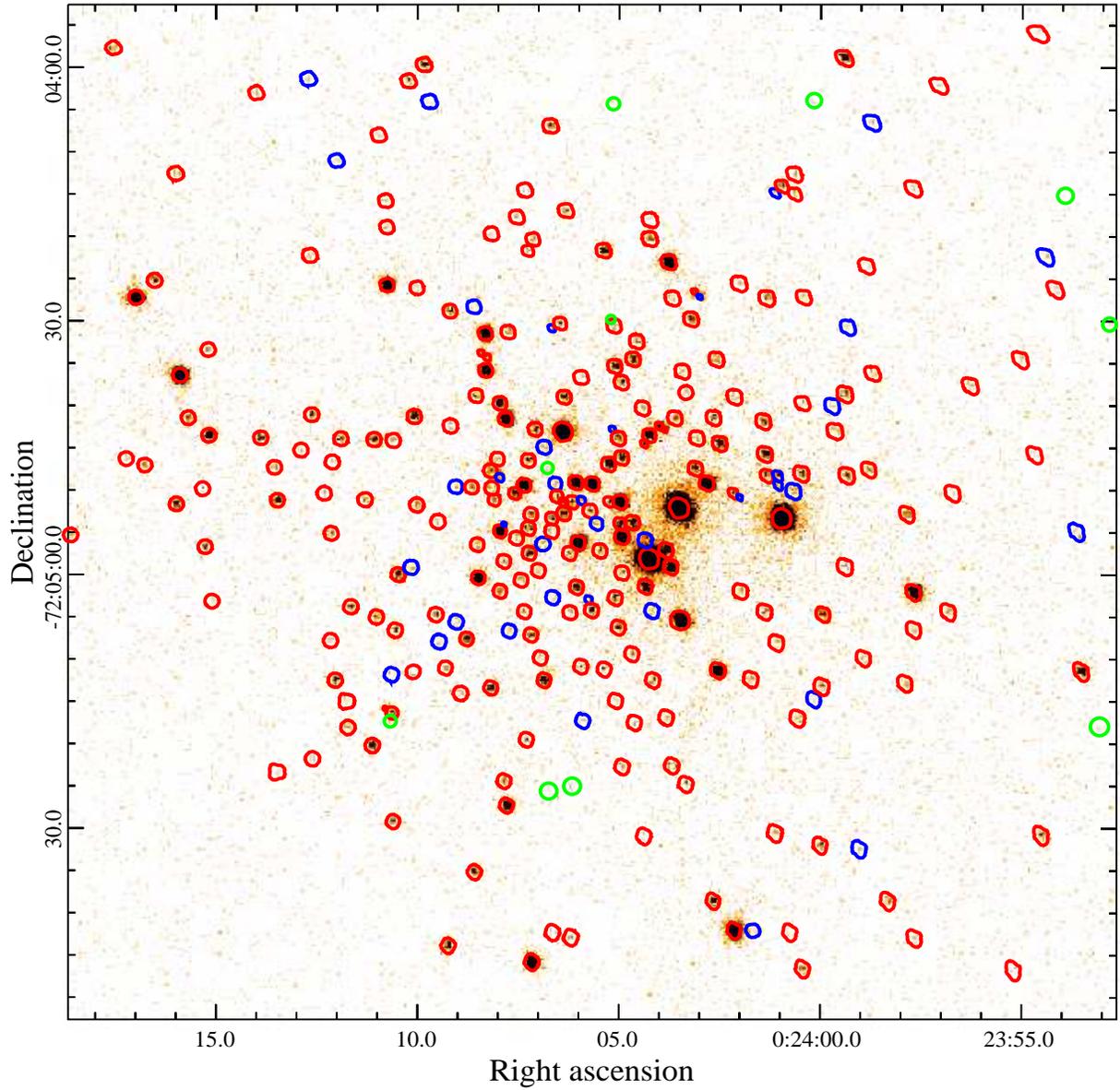}
\linespread{-0.7}
\caption{
{\it Chandra} merged image of the central $2\arcmin \times 2\arcmin$ region of 47 Tuc. The image was binned to $0.25\arcsec$ per pixel. The color-coded symbols denote the different types of sources, as in Figure-\ref{fig:rawimage}.
\label{fig:rgbimage}}
\end{figure*}

To obtain a reliable source list, we utilized AE to filter and refine the candidate source list, which was developed for analyzing multiple overlapping {\it Chandra} ACIS observations \citep{broos2010}. By analyzing the level-2 event files of each observation, 
AE builds source and background regions with local PSF contours\footnote{Two sets of regions can be constructed with AE in the photometric analysis of point sources: The contours of the local PSF, which is defined with a given enclosed counts fraction ($0<$ECF$<1$) and can be set as the source extraction regions; The ``mask regions", which completely cover the sources (the default mask region in AE is chosen to be 1.1 times a radius that encloses 99\% of the PSF) and can be used to construct the background regions.}, and creates the source and background spectra with appropriate extraction apertures. Then by merging the extractions of multiple observations into ``composite" data products (i.e., event files and spectra for source and background regions, PSF model, ARF and RMF, etc.), we are able to perform further analysis of source properties (such as source validation, position refining, photometry and spectral fitting, etc.) with AE \citep{broos2010}. 
Standard AE point-source extraction workflow usually involves many iterations of extracting, merging, pruning, and repositioning of sources in the candidate catalog\footnote{See Figure 8 of the AE User's Guide for details}, which is helpful to our evaluation of the significance of the X-ray sources in 47 Tuc. As illustrated in Figure-\ref{fig:rgbimage}, X-ray sources in the core of 47 Tuc are overcrowded, thus extraction of point sources is strongly affected by their neighbours. 

For these reasons, we adopted an iterative pruning strategy to examine the validation of each candidate source in 47 Tuc. Our source pruning processes is similar to the steps outlined in the validation  procedure\footnote{http://www2.astro.psu.edu/xray/docs/TARA/ae\_users\_guide/procedures/} presented by the AE authors: all candidate sources were extracted, merging with multiple extractions, pruning insignificant sources, repositioning and checking astrometry repeatedly, until a stable source list with refined positions was obtained. AE provides an important output parameter, the binomial no-source probability $P_{B}$, to evaluate the significance of a source. The evaluation of $P_{B}$ is based on the null hypothesis that a source does not exist in the source-extraction aperture, and the observed excess number of counts over background is purely due to background fluctuations \citep{weisskopf2007}. The formula to obtain $P_{B}$ is given by
\begin{equation}
P_{\rm B}(X\ge S)=\sum_{X=S}^{N}\frac{N!}{X!(N-X)!}p^X(1-p)^{N-X},
\end{equation}
where $S$ is the total number of counts in the source extraction region and $B$ is the total number of counts in the background extraction region; and $N$ is the sum of $S$ and $B$; $p=1/(1+BACKSCAL)$, with $BACKSCAL$ being the area ratio of the background and source-extraction regions. 
For each source, AE computed a $P_{B}$ value in each of the three bands, and we adopted the minimum of the three as
the final $P_{B}$ value for the source.
By adopting an invalid threshold value of $P_{B}< 1\times 10^{-3}$ for the candidate source list, we obtained a final stable catalogue with 537 X-ray sources for 47 Tuc.

As comparison, our final catalog of X-ray sources is much larger than that obtained by \citet{bhattacharya2017}, which contains 370 X-ray sources within a radius of $2.79\arcmin$ in 47 Tuc. 
Although our X-ray source searching region is larger (i.e. $R\sim$ 7.5\arcmin), we found that 61 of the new detections (blue sources in Figure-\ref{fig:rawimage} and \ref{fig:rgbimage}) are located within $R=2.79\arcmin$ in our catalog, while 28 sources (green sources in Figure-\ref{fig:rawimage} and \ref{fig:rgbimage}) in \citet{bhattacharya2017} are failed to be recovered in our catalog. Discrepancy may result from the different source detection and pruning strategies. 
By adopting a loose source-detection threshold in the {\it wavdetect} script, we allow more fainter X-ray sources to be detected, which ensure the source completeness (i.e., recovery number of real sources) of our catalog, but with a sacrifice to enroll some spurious sources.  
In order to optimize the balance between the source completeness and reliability (i.e., the fraction of potential spurious sources), we adopted a more strict binomial no-source probability threshold (i.e., $P_{B}< 1\times 10^{-3}$) in AE source pruning procedure, which is much less than the value ($P_{B}< 1\times 10^{-1}$) used by \citet{bhattacharya2017}. 

\begin{figure*}[htp]
\centering
\includegraphics[angle=0,origin=br,height=0.34\textheight, width=1.0\textwidth]{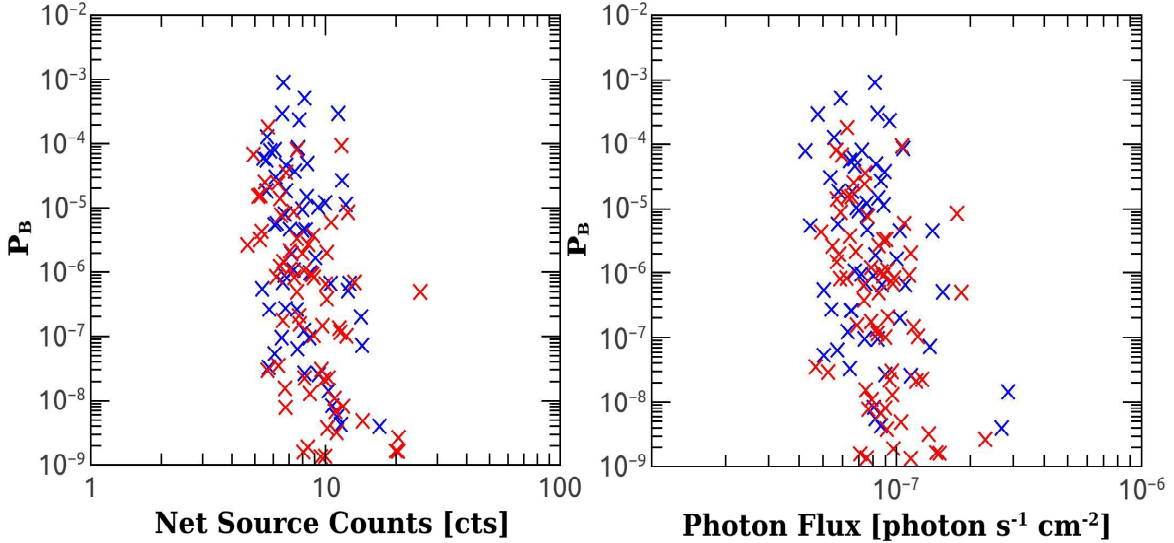}
\linespread{-0.7}
\caption{
Binomial no-source probability as a function of the 0.5-8 keV net source counts and photon flux. Only dim (i.e., $P_{B}>1.0\times 10^{-9}$) X-ray sources located within $R=2.79\arcmin$ have been plotted in this figure. More bright sources, with larger net source counts or photon flux, were not shown here due to the much lower value of $P_B$. Sources cross-identified in \citet{bhattacharya2017} and this work are marked as red crosses, while new detections of this work are shown as blue crosses.  
 \label{fig:pb}}
\end{figure*}

To compare the performance of our source detection and pruning strategies with that of \citet{bhattacharya2017}, we plot $P_{B}$ as a function of net source counts and photon flux in Figure-\ref{fig:pb}. Obviously, the new detections (i.e., blue sources in Figure-\ref{fig:rawimage} and \ref{fig:rgbimage}) of our work are uniformly mixed with the validated detections (i.e., red sources that have been cross-identified in \citealp{bhattacharya2017} and this work), which supports the better source completeness in our catalog.
For green sources in Figure-\ref{fig:rawimage} and \ref{fig:rgbimage}, many of them were found to have few photons ($\lesssim 5$) and much higher value of binomial no-source probability ($1.5\times 10^{-2} \lesssim P_{B}\lesssim 1.0\times 10^{-1}$). They were labelled as ``$c$" and were thought to be marginal detections in the catalog of \citet{bhattacharya2017}. With a more stringent pruning $P_{B}$ threshold (i.e., $P_{B}< 1\times 10^{-3}$), these sources have been removed from our catalog.
We note that the choice of the $P_{B}$ threshold is an empirical decision, which may range from $10^{-2}$ (AE default) to $10^{-3}$ in literature.

\subsection{Source Properties}
Following source pruning and position refinement, we utilized AE to extract final source and background spectra for the X-ray sources.
The default AE extraction aperture was defined to enclose $\sim 90\%$ (evaluated at 1.5 keV) of the PSF power, which may lead to overlapped extraction regions in the dense core of 47 Tuc.
By reducing the aperture sizes iteratively and calculating the energy-dependent photometry correction factors for the individual sources, AE provides a sophisticated method to perform source extraction for crowding environment.
In our AE usage, we adopted the default AE polygonal region as the source extraction regions for most sources in 47 Tuc, and reduced the extraction regions to enclose $\approx 40\%-90\%$ of the local PSF power for sources with close neighbours.
For the background extractions, we adopted the AE ``BETTER\_BACKGROUNDS" algorithm to build the background regions. This algorithm models the spatial distributions of flux for the source of interest and its neighboring sources using unmasked data, and then computes local background counts within background regions that subtract contributions from the source and its neighboring sources. By setting a minimum number of 100 counts for each merged background spectrum to ensure the photometric accuracy, we obtained the background regions of each source through the ``ADJUST\_BACKSCAL" stage in AE.

Spectral extractions are first performed independently for each source and each observation, then using AE we have merged the extraction results to construct composite products (i.e., event lists, spectra, light curves, response matrices and effective area files) for each source through the “MERGE\_OBSERVATIONS” procedure. 
Aperture-corrected net source counts are derived in 0.5--2 keV (soft), 2--8 keV (hard) and 0.5--8 keV (full), respectively. About $45\%$ (i.e., 244/537) of the X-ray sources have net counts greater than $\sim 30$ in the full band, and are available for spectral analysis with the AE automated spectral fitting script. We modeled the spectra of these sources with an absorbed power-law spectrum. In all cases the neutral hydrogen column density ($N_{H}$) is constrained to no less than $2.3 \times 10^{20} {\rm \ cm^{-2}}$, calculated from the color excess E(B--V) of 47 Tuc.
For the remaining sources, their net counts are less than $\sim 30$ and the X-ray spectra are therefore poorly constrained. We converted their net count rates into unabsorbed fluxes, by using the AE-generated merged spectral response files and assuming a power-law model with fixed photon-index ($\Gamma=1.4$) and column density ($N_{H}=2.3 \times 10^{20} {\rm \ cm^{-2}}$). Taking a distance of 4.02 kpc for 47 Tuc \citep{mcLaughlin2006}, we converted the flux of each X-ray source into unabsorbed luminosity in soft, hard and full bands, respectively.
\begin{figure*}[htp]
\centering
\begin{minipage}[!htbp]{1.0\textwidth}
\leftline{\includegraphics[angle=0,origin=br,height=0.33\textheight, width=0.5\textwidth]{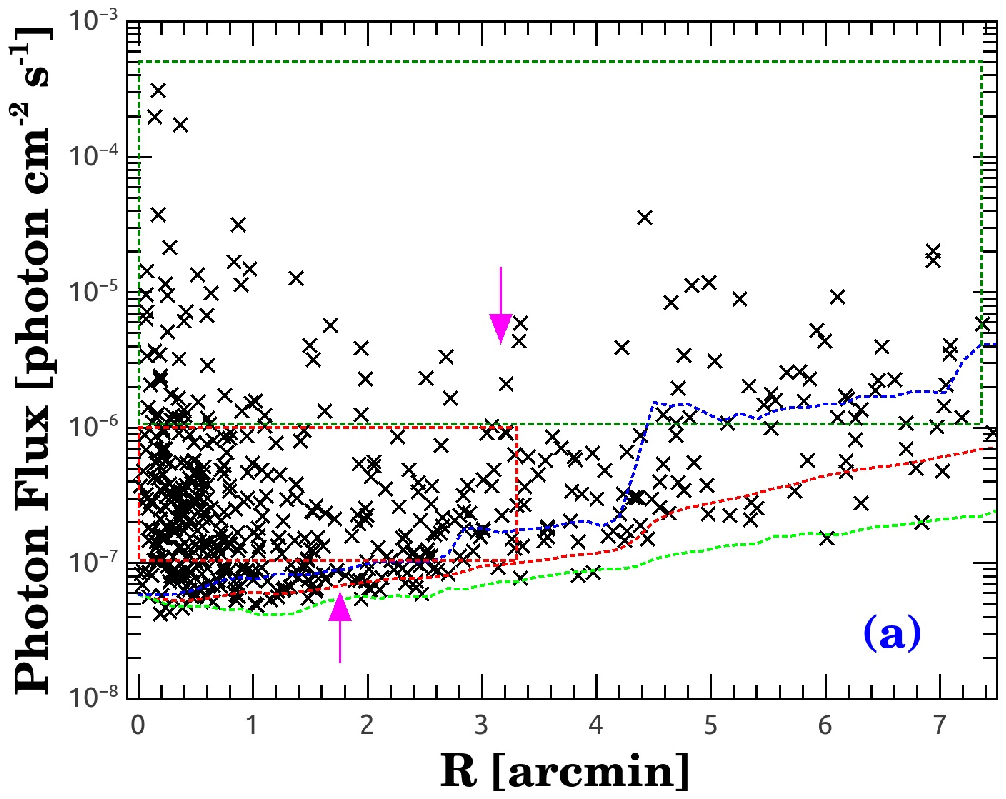}
\includegraphics[angle=0,origin=br,height=0.33\textheight, width=0.5\textwidth]{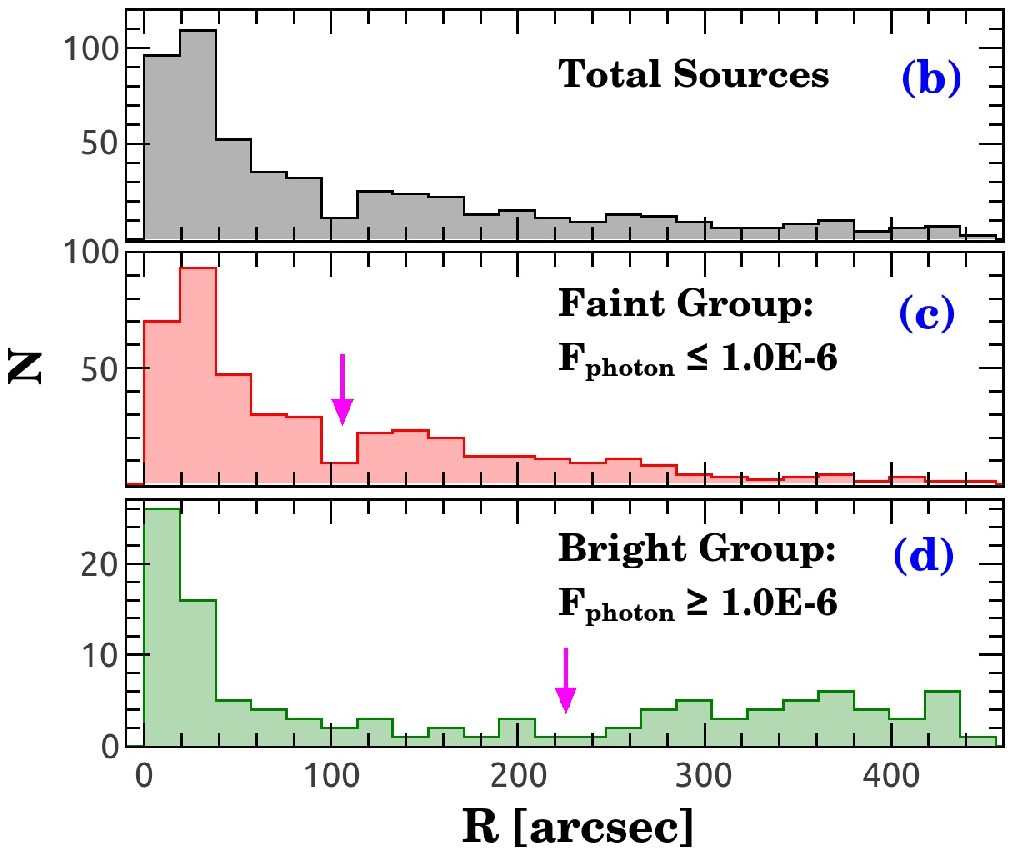}}
\linespread{0.7}
\caption{(a): 0.5-8 keV photon flux as a function of projected distance from the cluster center. The color-coded lines represent the median (red), minimum (green) and maximum (blue) limiting detection count rate at corresponding radial bins, respectively. The total source samples ((a) and (b)) are classified into the faint (c) and bright group (d) of X-ray sources separately, and their surface density distributions are displayed in Figure-\ref{fig:surfd}. Samples selected by the dotted boxes in (a) are shown in Figrue-\ref{fig:distribution} and Figure-\ref{fig:fit}, which have the advantage of unbiased detection. The two distribution dips are marked by magenta arrows. \label{fig:pflux}}
\end{minipage}
\end{figure*} 

Finally, we collated the source extraction and spectral fitting results into a main X-ray source catalog for 47 Tuc. The X-ray sources are sorted by their Right Ascension. We calculated their distance to cluster center using the optical central coordinate $\alpha =00^{h}24^{m}05^{s}.67$ and $\delta =-72\arcdeg 04\arcmin 52.62\arcsec$ \citep{mcLaughlin2006}. 
Our catalog contains 22 columns of information for the 537 X-ray sources (Table-\ref{tab:mcat}), and the full
table is available in the electronic edition of the Journal.

\subsection{Sensitivity}
As shown in Figure-\ref{fig:rawimage}, the merged images have the deepest exposure near the cluster core and much shallow exposure at the cluster halo, which leads to significant variation of detection sensitivity (i.e., the minimum flux at which a source would be detected) across the searching region.
In order to estimate the sensitivity of each point in the survey field, we created background and sensitivity maps following the procedures described by \citet{xue2011} and \citet{luo2017} in {\it Chandra} deep field-south survey.
Briefly, we first masked out the main catalog sources from the full band merged image using circular regions with mask radii ($r_{msk}$) provided by AE. We then filled in the masked region for each source with the background counts from a local annulus with inner to outer radii of $r_{msk}-2.5r_{msk}$. 
Due to the crowding of X-ray sources, the default annuli for background counts are strongly overlapped with neighbouring sources in the core of 47 Tuc. 
Therefore, for X-ray sources located in $R\leq20\arcsec$ and $20\arcsec<R\leq40\arcsec$, we filled in them with background counts from an uniform background region separately. This uniform background region was defined by subtracting the mask regions of detected X-ray sources within the circle region with $R\leq 20\arcsec$ and the annulus region with $20\arcsec\leq R\leq40\arcsec$, respectively. 
  
The resulted source-free background map was then used to calculate the limiting sensitivity map, which is the flux limit required for a source to be selected by our $P_{B}$ criterion.
To follow the behavior of AE in photometry extraction of the main-catalog sources, we defined for each pixel a circular source extraction aperture with the local $90\%$ PSF ECF radii. 
Due to the off-axis effects, the value of $BACKSCAL$ (i.e., area ratio of the source to background extraction regions) is depends on the off-axis angles in AE. Therefore, for a given pixel in the survey field, we also computed its off-axis angle $\theta_{p}$, and set the value of $BACKSCAL$ to the maximum value of the main-catalog sources that are located in an annulus with the inner/outer radius being $\theta_{p}-0.25\arcmin/\theta_{p}+0.25\arcmin$ (note that the adopted maximum $BACKSCAL$ value corresponds to the highest sensitivity). 
With the defined source extraction aperture, $BACKSCAL$ and background map, we then calculated for each pixel the background counts ($B$). 
The detection sensitivity ($S$), which is the minimum number of source counts required for a detection, can be obtained by solving Equation (1) with survey $P_{B}$ threshold value of $1\times 10^{-3}$.
Finally, We computed for each pixel the limiting detection count rates with the exposure map, and converted it into limiting fluxes by assuming an absorption power-law model with photon-index $\Gamma=1.4$ and column density $N_{H}=2.3 \times 10^{20} {\rm \ cm^{-2}}$.

The limiting detection count rates of each pixel are shown as a function of projected distance from the cluster center in Figure-\ref{fig:pflux}(a). The median, minimum and maximum value of detection limit at each radial bin is represented as red, green and blue dotted lines, respectively. 
For comparison, we also plot the photon flux of the 537 X-ray sources as diagonal crosses in Figure-\ref{fig:pflux} (a). 
As the figure suggests, the merged images have a deepest and relatively flat limiting sensitivity ($F_{p}\sim6\times 10^{-8} {\rm \ photon\ cm^{-2} s^{-1}}$) within $R\lesssim 2.8\arcmin$. 
At larger cluster radii, the defined survey field is poorly covered by the {\it Chandra} (subarry) ACIS-S Field of Views (see also Figure 1), thus, the median of the sensitivity curve decreases significantly with increasing $R$. However, almost all of the main-catalog X-ray sources are located above the minimum sensitivity line.

\begin{figure*}[htp]
\centering
\begin{minipage}[!htbp]{1.0\textwidth}
\leftline{\includegraphics[angle=0,origin=br,height=0.4\textheight, width=0.5\textwidth]{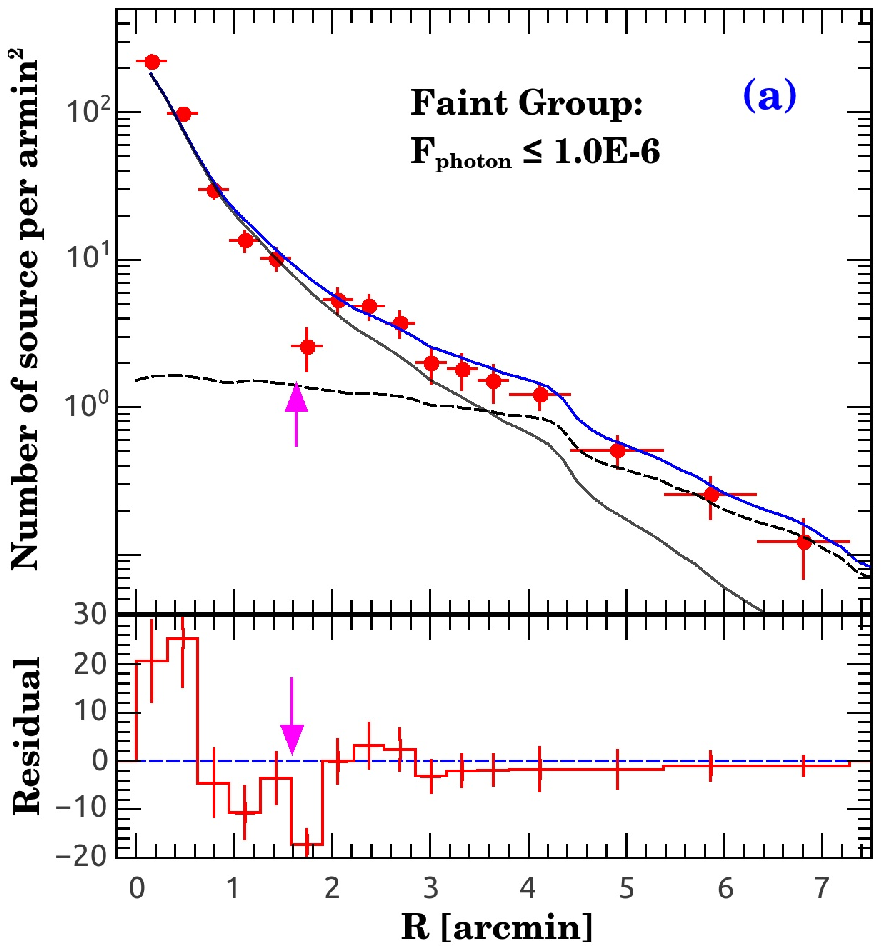}
\includegraphics[angle=0,origin=br,height=0.4\textheight, width=0.5\textwidth]{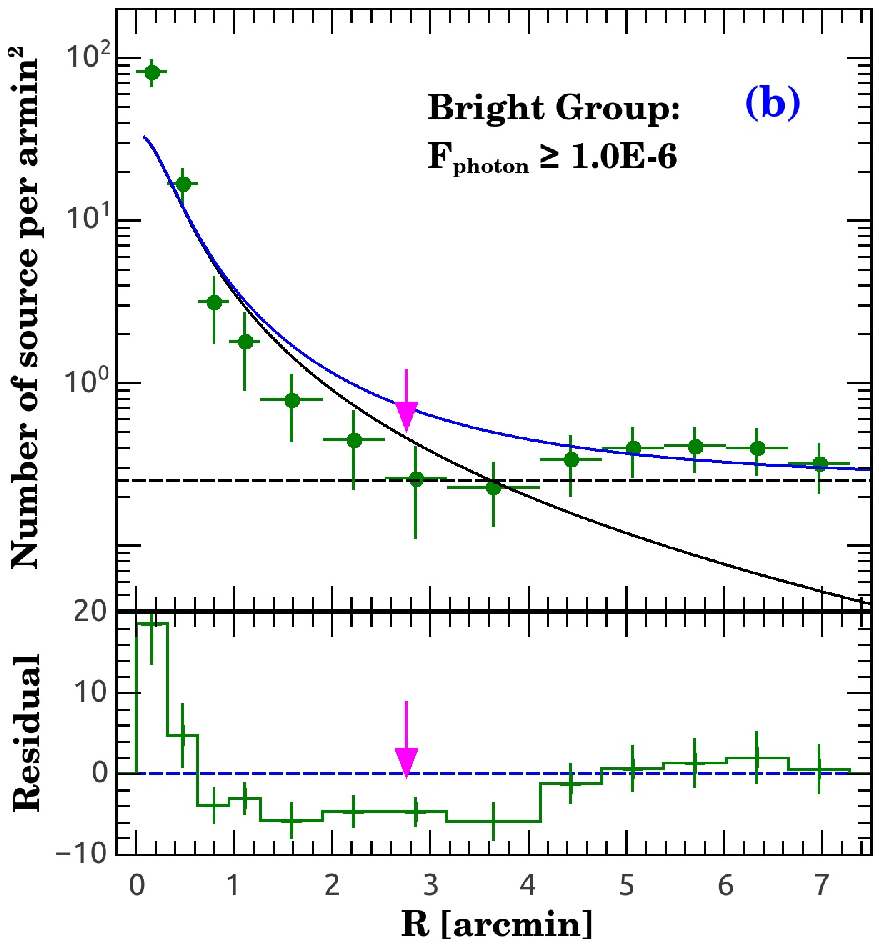}}
\linespread{0.7}
\caption{Radial surface density distribution of the faint (a) and bright (b) groups of X-ray sources in 47 Tuc. The red and olive dots donate the observed X-ray sources. The black dashed curves represent the CXB contributions. Due to the substantial decrease of the effective exposure, the CXB component in (a) drops significantly at large radius (i.e., beyond $\sim 4\arcmin$). The black solid lines mark the profile of stellar density convolving with the sensitivity function, which was calculated with the best-fitting King model of 47 Tuc and normalized to match the total number of GC X-ray sources in each group. The solid blue lines are the sum of CXB and the King model components.
The lower panels show the residual of the observed X-ray sources ($N_{X}$) after subtracting the model predicted X-ray sources ($N_{K}+N_{B}$). X-ray sources are over-abundant in cluster center and scarce at the distribution dips (magenta arrows) with respect to the distribution of cluster light. \label{fig:surfd}}
\end{minipage}
\end{figure*}

\section{Analysis: Source Radial Distribution}

Despite the sensitivity variation across the survey field, we found that X-ray sources in 47 Tuc can be divided into several groups according to their locations in Figure-\ref{fig:pflux}(a). 
Specifically, the radial distribution of the bright X-ray sources (i.e., sources with photon flux $F_{p}\gtrsim 1\times 10^{-6} {\rm\ photon\ cm^{-2}\,s^{-1}}$ or luminosity $L_{X}\gtrsim 5.0\times 10^{30} {\rm\ erg\,s^{-1}}$) is bimodal, with two peaks located at the cluster center and $R\sim 6\arcmin$, respectively. 
A broad dip exists between $1\arcmin\lesssim R\lesssim 4\arcmin$ (marked by magenta down arrow), and a narrow dip is evident around $1.5\arcmin\lesssim R\lesssim 1.8\arcmin$ (magenta up arrow) for the faint X-ray sources (i.e., $F_{p}\lesssim 1\times 10^{-6} {\rm\ photon\ cm^{-2}\,s^{-1}}$ or $L_{X}\lesssim 5.0\times 10^{30} {\rm\ erg\,s^{-1}}$). 
These distribution dips can also be clearly identified from Figure-\ref{fig:pflux}(b), \ref{fig:pflux}(c) and \ref{fig:pflux}(d), where the radial distributions of total, bright and faint X-ray sources are presented as black, red and olive histograms, respectively. 
Since the 47 Tuc is very close (with an angular distance of $\sim2.3^{\circ}$) to the Small Magellanic Cloud (SMC), some of the X-ray sources could be associated with the SMC. We check this possibility by examining the azimuthal distribution of sources located within $3\arcmin \lesssim R \lesssim 7.5\arcmin$, and found that there is no significant excess of X-ray sources toward the direction of the SMC.
We also examined whether these distribution dips are created by the coverage effects (i.e., CCD gaps or variation of detection sensitivity) of the merged observations, and found that such effects are negligible (Figure-\ref{fig:rawimage}).

However, classification of X-ray sources in Figure-\ref{fig:pflux} is challenging, due to the contamination of cosmic X-ray background (CXB). Assuming a uniform spatial distribution, the CXB contamination will become dominant outside the cluster core since the survey area grows rapidly with increasing $R$. 
To estimate the contamination of CXB sources, we plot the X-ray source surface density as a function of $R$ in Figure-\ref{fig:surfd}. The observed X-ray sources are displayed as red and olive dots, while the CXB contributions are shown as black dashed lines. 
Here we adopted the ${\rm log}N-{\rm log}S$ relations determined by \citet{kim2007} to estimate the contribution of CXB sources. The cumulative number count of CXB sources of a limiting sensitivity flux $S$ can be estimated with function
\begin{equation}
{N(>S)}=2433(S/10^{-15})^{-0.64}-186\ {\rm deg^{-2}}.
\end{equation}
This function is derived from Equation (5) of \citet{kim2007}, by assuming a photon index of $\Gamma=1.4$ for the {\it Chandra} ACIS observations in 0.5-8 keV band. 

Figure-\ref{fig:surfd} suggests the excess of X-ray sources over the CXB can be reasonably accounted for by the GC X-ray sources. For the faint group of X-ray sources, the excess ($\sim336$ sources) drops rapidly with increasing $R$ and becomes insignificant at $R\sim 6.0\arcmin$. Beyond $R\sim 6.0\arcmin$, the predicted CXB contribution matches well with the observed surface density profile. 
For the bright group of X-ray sources, the excess ($\sim 70$ sources) over the CXB is evident within $R\lesssim 2.5\arcmin$ and $4\arcmin \lesssim R\lesssim 7.5\arcmin$, and becomes negligible at $2.5\arcmin \lesssim R\lesssim 4\arcmin$. Apparently, even when the CXB sources are taken into account, the two distribution dips are still evident in Figure-\ref{fig:surfd}.

To quantitatively evaluate the significance of the distribution dips, we modeled the radial distribution of GC X-ray sources with the stellar density profile of 47 Tuc.  
Through out this work, we utilized the King model \citep{king1962} to calculate the radial distribution of normal stars in 47 Tuc. The cluster concentration ($c=2.01$) and core radius ($r_{c}=20.84\arcsec$) are adopted from \citet{mcLaughlin2006}, which were determined using stars with mass of $0.85M_{\odot}\lesssim M_{\ast} \lesssim 0.9 M_{\odot}$. 
We then convolved the King model profile with the sensitivity function, and normalized the model profile to match the total number of GC X-ray sources in each group.
The King model profiles are plotted as black solid lines in Figure-\ref{fig:surfd}, while the sum of the CXB and King model components are represented as blue solid lines.
We defined the significance of the distribution dip as $(N_{K}+N_{B}-N_{X})/\sqrt{N_{K}^{2}\sigma_{K}^{2}+N_{B}^{2}\sigma_{c}^{2}+N_{X}^{2}\sigma_{P}^{2}}$. 
Here $N_{K}$ is the number of X-ray sources predicted by the King model, $\sigma_{K}$ is the fitting error for the King model, $\sigma_{P}=1/\sqrt{N_{X}}$ is the the Poisson error for the observed number of sources ($N_{X}$), and $\sigma_{c}^{2}$ is the variance of CXB sources ($N_{B}$). By adopting a nominal error of $\sigma_{K}=5\%$ for the King model estimation and a CXB variance of $\sigma_{c}^{2}=2.25\%$ for the {\it Chandra} surveyed field (with an area of $\sim 0.05 \rm \, deg^{2}$) in 47 Tuc \citep{moretti2009}, we adjust the locations and widths of the distribution dips, and found a maximum significance of $\sim7.3\sigma$ and $\sim6.3\sigma$ for the bright (with $119\arcsec \leqslant R\leqslant 253\arcsec$, $N_{K}=16.5$, $N_{B}=10.9$ and $N_{X}=7$) and faint (with $99\arcsec \leqslant R\leqslant 113\arcsec$, $N_{K}=16.9$, $N_{B}=3.5$ and $N_{X}=4$) distribution dips, respectively.

Compared with BSS, the split of X-ray sources into two sub-populations in radial direction suggests that they may have experienced similar mass segregation effect. To test this hypothesis, we set normal stars as reference and compare their cumulative radial distribution with that of the X-ray sources in 47 Tuc.
Due to the variation of detection sensitivity across the surveying field, we need to ensure that the selected sample of the X-ray sources are observationally unbiased. To maximize the sizes of source samples and minimize the CXB contamination, we set a lower photon flux limit of $\sim 1\times 10^{-7} {\rm \ photon\ cm^{-2} s^{-1}}$ for the faint group of X-ray sources, and confined the source selection region as $R \lesssim 200\arcsec$ (i.e., red dotted box in Figure-\ref{fig:pflux}(a)). The final sample contains 250 X-ray sources. About 35 of them are CXB sources. For the bright group of X-ray sources, 113 sources are located within $R \lesssim 450\arcsec$ (olive dotted box in Figure-\ref{fig:pflux}(a)), where $\sim 43$ sources are from the CXB.

Assuming that the CXB sources are uniformly distributed across the survey field, we corrected the cumulative distribution of X-ray source samples for CXB contamination, following the Monte Carlo procedure described in \citet{grindlay2002}. A total of 1000 bootstrap resamplings were generated for each group of X-ray sources. For each of the bootstrap sample, we simulated the CXB sources using a Poisson distribution with a mean number of 35 (or 43). The CXB sources were set to have a uniform spatial distribution, and the closest actual sources to these positions were removed from the bootstrap sample. The average of the 1000 background-corrected sample distributions was then adopted as the best estimate for the corrected cumulative distribution.

\begin{figure*}[htp]
\leftline{
\includegraphics[angle=0,origin=br,height=0.45\textheight, width=1.05\textwidth]{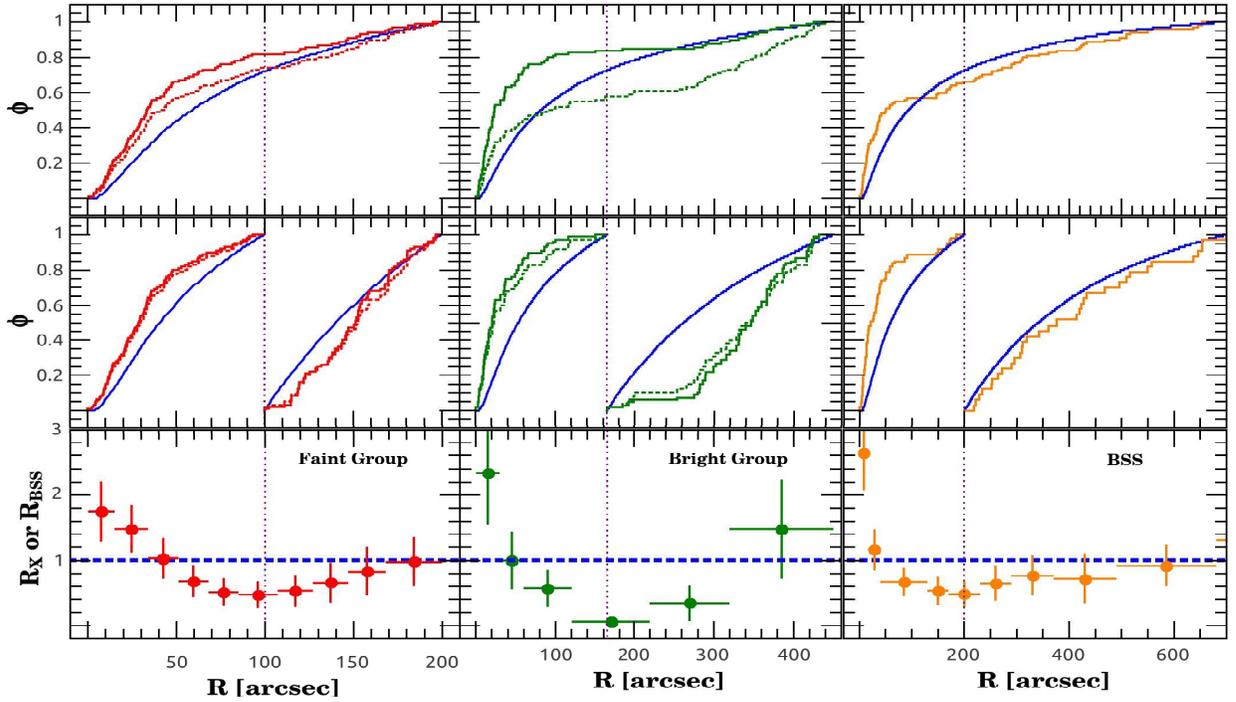}}
\linespread{0.7}
\caption{Top panels: Cumulative distributions of X-ray sources and BSS as a function $R$ in 47 Tuc. The color coded step lines represent the faint (red) and bright (olive) groups of X-ray sources, and BSS (orange), respectively, while the blue solid lines represent the distributions of normal stars predicted by the King model. The observed distributions of X-ray sources are shown in dotted lines and the CXB-corrected distributions were shown as solid lines. The vetical dotted lines mark the locations of the distribution dips. Middle panels: Same as the top panels, but the selected sources were divided into two sub-populations according to their distibution dips. Bottom panels: Specific frequency of X-ray sources and BSS as a function of the distance from the cluster center for each of the concentric annuli defined in Table-\ref{tbl:spec_freq}. The horizontal dashed line with specific frequence $R_{X}=1$ corresponds to a uniform distribution relatively to the integrated light of the cluster. Data of BSS were adopted from \citet{ferraro2004}. \label{fig:distribution}}
\end{figure*}

Following the procedures presented by \citet{ferraro1993} and \citet{ferraro2004}, we first plot the cumulative distribution of X-ray sources as a function of $R$ in the upper panels of Figure-\ref{fig:distribution}. 
The faint and bright group of X-ray sources are shown as red and olive step lines, while the cumulative distributions of the reference normal stars are displayed as blue lines.
Apparently, the X-ray sources are more concentrated than the normal stars in the central region, and less concentrated in the outer region.
This feature is better illustrated in the middle panels of Figure-\ref{fig:distribution}, in which the two sub-populations of each group of X-ray sources are shown separately. These distribution profiles are similar to that of BSS (right panels of Figure-\ref{fig:distribution}), which suggests a universal mass segregation effect for X-ray sources and BSS in 47 Tuc.

As suggested by \citet{ferraro1993}, the mass segregation effect of X-ray sources in GCs can also be studied by using the number of X-ray sources normalized to the integrated cluster light. In this case the survey area is divided into a set of concentric annuli, and the number of X-ray sources ($N_{X}$) in each annulus is counted and then normalized to the total light emitted in that annulus. In order to correct the contamination (i.e., $N_{B}$) of CXB sources, We modified the specific frequency equation of \citet{ferraro1993} to
\begin{equation}
R_{X}=\frac{(N_{i}/N_{t})}{(L_{i}/L_{t})},
\end{equation}
where $N_{i}=N_{X}-N_{B}$ ($N_{t}=\Sigma_{i} N_{i}$) is the background subtracted number of X-ray sources within each annulus (survey area), and $L_{i}$ ($L_{t}$) is the corresponding V-band luminosity from the annulus (survey area). Following the method used by \citet{ferraro1993}, we utilized the King model to estimate the integrated light ratio (i.e., $L_{i}/L_{t}$) for each annulus, and assigned a nominal error of 5\% for the King model estimation. By assuming a Poisson error ($\delta N_{i}=\sqrt{N_{i}}$) for X-ray source counts, we calculated the error of $R_{X}$ with the formula
\begin{equation}
\sigma_{R}=(\beta\sigma_{b}+\sigma_{a})/b,
\end{equation}
where $\beta=a/b$, $\sigma_{a}$ and $\sigma_{b}$ is the error of $a$ and $b$, respectively. We listed the data and results for each annulus in Table-\ref{tbl:spec_freq}.  

In the bottom panels of Figure-\ref{fig:distribution}, we plot the specific frequency of X-ray sources as a function of $R$. The faint and bright group of X-ray sources are presented as red and olive dots, respectively. For comparison, we also plot a blue dashed horizontal line (i.e., $R_{X}=1$) in the bottom panels of Figure-\ref{fig:distribution}, which describes a uniform distribution of X-ray sources with respect to the integrated light of cluster. As shown in the figure, the specific frequency of X-ray source ($R_{X}$) reaches its maximum at the cluster center, decreases to the minimum at the distribution dips and then rises again. Compared with the dashed horizontal line, the distribution profile of $R_{X}$ suggests that the mass segregation is efficient within the distribution dips and many X-ray sources have been drifted into the cluster core, leading to an over-abundance of X-ray sources at the center of 47 Tuc. Beyond the distribution dips, mass segregation is inefficient, and lots of X-ray sources remain a large orbit from the cluster center. As a result, an annulus region devoid of X-ray sources is formed at the distribution dips.

For comparison, we also plot the cumulative and specific frequency distributions of BSS (orange lines and symbols) in the right panels of Figure-\ref{fig:distribution}. The data of BSS in 47 Tuc were adopted from \citet{ferraro2004}, and only BSS with $0<R<700\arcsec$ have been illustrated in Figure-\ref{fig:distribution}. We note that the locations and widths (i.e., range of $R$ with $R_{X}\lesssim 1.0$) of the distribution dips are $R \sim 100\arcsec$ and $\Delta R \sim 100\arcsec$ for the faint group of X-ray sources, $R \sim 170\arcsec$ and $\Delta R \sim 300\arcsec$ for the bright group of X-ray sources, $R \sim 200\arcsec$ and $\Delta R \sim 500\arcsec$ for the BSS, respectively.
 
\section{Discussion}
\subsection{Mass Segregation of X-ray Sources and BSS in 47 Tuc} 
As described by \citet{ferraro2012}, the two-body relaxation is the main process that drives more massive objects (such as binaries and BSS) sedimentation to the cluster center, and modifies an initially flat BSS radial distribution into a bimodal shape.
Due to the highest stellar density, mass segregation is thought to take place at cluster center first.
As time goes on, heavy objects orbiting at larger radii are expected to drift toward the core. As a consequence, the region devoid of these heavy objects will propagate outward progressively, and the two-body relaxation timescale at the distribution dip is roughly scaled to the dynamical age of the cluster.
The two-body relaxation timescale in GCs can be expressed as \citep{binney2008}:
\begin{equation}
t_{relax}\propto\frac{{\sigma}^{3}(r)}{G^{2}M{\rho}(r) ln\Lambda},
\end{equation}
where $\rho(r)$ and $\sigma(r)$ is the profile of stellar density and stellar velocity dispersion, respectively, $G$ is the gravitational constant, $ln\Lambda$ is the Coulomb logarithm, and $M$ is the average mass of the heavy objects.
In GCs, the stellar density profile ($\rho(r)$) decreases dramatically outside the cluster core, which leads to a steep increase of the two-body relaxation timescale from the cluster center to the halo\footnote{For example, the typical relaxation timescale in cluster core ($t_{rc}$) is about 1-3 orders of magnitude lower than the Hubble timescale ($t_{H}$), while typical relaxation timescale at half-light radius ($t_{rh}$) is slightly less than or comparable to $t_{H}$ \citep{harris1996}.}. For 47 Tuc, the two-body relaxation time is $t_{rh}\approx 3.55\, \rm Gyr$ at the half-light radius $r_h=3.17\arcmin$ \citep{harris1996}, which is much smaller than the age ($t=13.06 \, \rm Gyr$, \citealp{forbes2010}) of this cluster. Therefore, significant mass segregation effect is expected to in this cluster.

In Figure-\ref{fig:distribution}, we show that the distribution dips of X-ray sources and BBS are different in 47 Tuc. Considering that initial conditions and cluster environment (i.e., $\rho(r)$, $\sigma(r)$ and $ln\Lambda$) should be same for all of these objects, the mass of heavy objects becomes the only factor that affects their distribution dips in this cluster.
According to Equation (5), massive objects drift to the cluster center faster than the less massive ones, and their radial distribution dip propagates outward further. If this is the case, Figure-\ref{fig:distribution} suggests that BSS in 47 Tuc will be more massive than the X-ray sources, and the bright X-ray sources should be more massive than the faint X-ray sources. Thus, an estimation of the average mass of each group of heavy objects is helpful to clarify this issue. 

In GCs, radial distribution of heavy objects can be used to estimate their average mass. For example, by analyzing the distribution of the projected radial distance of LMXBs from cluster centers, \citet{grindlay1984} estimated the typical mass of LMXBs to be $\sim 1.5^{+0.4}_{-0.6}\ M_{\odot}$. 
Their method assumes that the distributions of LMXBs and normal stars are in thermal equilibrium. Therefore, LMXBs will be more centrally concentrated than that of the normal stars, which allows an estimation of the average mass ratio ($q=M_{X}/M_{\ast}$) between these two groups of objects. 
\citet{grindlay2002} presented a maximum likelihood procedure for fitting ``generalized King models'' to the radial distribution of heavy objects in clusters. In this model, the projected surface density profile of heavy objects takes the form (see also \citealp{heinke2005})
\begin{equation}
S(r)=S_{0}{\biggl[1+\biggl(\frac{r}{r_{c}}\biggr)^{2}\biggr]}^{(1-3q)/2},
\end{equation}
where $r_{c}$ is the cluster core radius, $S_{0}$ is the normalization constant, and $q=M_{X}/M_{\ast}$ is the average mass ratio of heavy objects over the reference stars, respectively. 

As suggested previously, X-ray sources located within the distribution dips are dynamically relaxed, hence energy equilibrium can be established between them and the normal stars. This is the presumption for estimating the mass of heavy objects with the radial distributions. Beyond the distribution dips, the local two-body relaxation timescale of each group of heavy objects could be much larger than the cluster dynamical age, thus no energy equilibrium can be established and we have to exclude these sources from our fitting samples. Finally, in order to better illustrate and compare the radial distribution of each group of heavy objects, we further constraint the ``generalized King models'' fitting region as $R<100\arcsec$ for all selected sources in Figure-\ref{fig:distribution}.

\begin{figure}[htp]
\leftline{\includegraphics[angle=0,origin=br,height=0.33\textheight, width=0.5\textwidth]{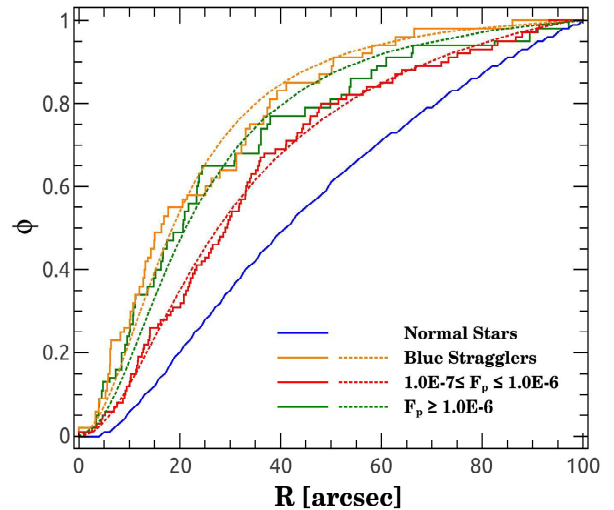}}
\linespread{0.7}
\caption{Cumulative radial distributions of each group of heavy objects in 47 Tuc. For each group, the actual distribution (solid lines) and maximum-likelihood fit (dashed lines) are shown. Note that the BSS (orange) shows the highest degree of central concentration, followed by the bright (olive) X-ray sources and the faint (red) X-ray sources. The reference normal stars are shown as green lines. \label{fig:fit}}
\end{figure}

In Figure-\ref{fig:fit}, we plot the CXB-corrected radial distribution of each group of heavy objects as step lines. Among all the considered heavy objects, BSS (orange) shows the highest degree of concentration, followed by the bright (olive) and faint (red) group of X-ray sources. 
We applied the K-S test to these distributions to check the statistical significance of the differences. 
The test yields that the BBS and bright X-ray sources are different at the $15.5\%$ confidence level, while the bright and fain X-ray sources, BSS and faint X-ray sources are different at $97.9\%$ and $99.3\%$ confidence level, respectively.  
The distributions of BSS and bright X-ray sources are similar, but they belong to different populations in GCs, thus the lower probability of statistical difference may result from a similar average mass between these two types of objects.
We fit the these distributions (shown as dashed lines) using a maximum-likelihood algorithm, which gives $q=1.73\pm0.19$ for BSS, $q=1.65\pm0.17$ for the bright X-ray sources, and $q=1.33\pm0.06$ for the faint X-ray sources, respectively.
Taking an average mass of $0.875\pm0.025\ M_{\odot}$ for the reference normal stars \citep{mcLaughlin2006}, we derived an average mass of $1.51\pm0.17\ M_{\odot}$, $1.44\pm0.15\ M_{\odot}$ and $1.16\pm0.06\ M_{\odot}$ for the BSS, bright and faint X-ray sources, respectively. Here, the fitting confidence is at 1$\sigma$ level.

Our results for the BSS average mass are consistent with the those reported in literatures, where the mass of BSS in 47 Tuc ranges from $0.99 M_{\odot}$ \citep{de2005} to $1.7 M_{\odot}$ \citep{shara1997,gilliland1998}. 
By comparing the velocity dispersion profiles of BSS and main-sequence turn-off stars in GCs, \citet{baldwin2016} estimated an average mass of $1.7^{+0.56}_{-0.35} M_{\odot}$ for BSS in 47 Tuc, which is consistent within errors with our fit results in Figure-\ref{fig:fit}.
For X-ray sources, their mass depends on which types of object they belong to. 
Quiescent LMXBs in GCs are thought to have an average mass of $1.5^{+0.3}_{-0.2} M_{\odot}$ \citep{heinke2003}, this is consistent with the derived mass of luminous LMXBs in \citet{grindlay1984}. 
For the MSPs, CVs and ABs identified in 47 Tuc, a similar ``generalized King models'' fitting estimated that they have an average mass of $1.47\pm0.19 M_{\odot}$, $1.31\pm0.22 M_{\odot}$ and $0.99\pm0.13 M_{\odot}$, respectively \citep{heinke2005}. 
More importantly, \citet{heinke2005} found that the majority of bright X-ray sources in 47 Tuc are either CVs or quiescent LMXBs, which is much luminous than most of the ABs (see Figure 9 and 10 of \citet{heinke2005} for details). 
Besides, the mass of ABs in 47 Tuc was found to be correlated with their X-ray flux, and ABs brighter in X-rays are expected to be more massive \citep{edmonds2003b}.
These evidence may indicate that the bright X-ray sources are dominated by CVs, quiescent LMXBs or luminous ABs, thus their average mass is significantly larger than the faint group of X-ray sources, which are mainly dominated by faint ABs. 

\subsection{Formation of Weak X-ray Sources in GCs: Primordial Binary Evolution versus Dynamical Encounters} 

Since the discovery, the origin of weak X-ray sources (i.e., mainly CVs, ABs, MSPs and quiescent LMXBs, \citealp{heinke2010}) in GCs has received much attention. In the earlier studies, in which the exposure are shallow and only brighter X-ray sources (i.e., with $L_{X}\gtrsim 4\times 10^{31}\rm\,erg\,s^{-1}$, \citealp{pooley2003}) were detected, a strong correlation between the source counts and the stellar encounter rate has been found in a dozen of GCs, which are thought to be an evidence of the dynamical formation of X-ray sources in GCs \citep{pooley2003}. But when more GCs and more fainter X-ray sources (i.e., $L_{X}\lesssim 4\times 10^{31}\rm\,erg\,s^{-1}$) are included, this correlation become less significant\footnote{For example, see the fainter group of X-ray sources in the bottom panels of Figure-2 in \citet{pooley2006}, or the Section 3 of \citet{cheng2018a} for details.}, which suggests that the contribution of primordial binary channel can not be ignored in GCs \citep{pooley2006,cheng2018a}. The bimodal distribution of X-ray sources in 47 Tuc provides further support for the primordial binaries contribution of weak X-ray sources sources in GCs. Note that the distribution profiles of $R_{X}$ and $R_{BSS}$ are similar in Figure-\ref{fig:distribution}, which may suggest a universal origin for X-ray sources and BSS in GCs------namely, both X-ray sources and BSS are descendant from primordial binaries, and their radial distribution are consistent with the model of mass segregation of binaries in GCs. 
This prediction can further be tested by the finding of X-ray sources and BSS in the outskirts of 47 Tuc, where stellar density decreases dramatically and dynamical interactions are less important, thus mainly the primordial binary formation channel is responsible for the creation of these sources.

Nevertheless, the two-body relaxation is not the only process that affect the evolution of primordial binaries in GCs. As binaries migrate to the dense core of GCs, they may suffer frequent encounters with other stars, which can modify the properties of binaries greatly and abruptly. 
According to the Hills-Heggie law, the evolution of binaries in GCs depends on their bound energy with respect to the kinetic energy of intruding stars. Stellar encounters involving hard binaries tend to make them harder, whereas encounters involving soft binaries drive them softer and eventually to disruption \citep{hills1975,heggie1975,hut1993}. 
It is possible for MS binaries to exchange one of their primordial members with the intruding compact objects and lead to the formation of CVs or LMXBs in GCs. 
Alternatively, the interactions between binaries and other stars may accelerate the evolution of MS binaries and transform them into ABs, CVs or BSS.
As a result, even for the X-ray sources and BSS detected at cluster center, many of them are likely formed by the encounters of primordial binaries \citep{ferraro1997,mapelli2004,knigge2009,cheng2018b}.

Recently, observations have revealed two populations of CVs in GCs. The X-ray luminosity distribution of CVs in NGC 6397 \citep{cohn2010}, NGC 6752 \citep{lugger2017} and 47 Tuc \citep{sandoval2018} was found to be bimodal.
The bright CVs (i.e., young CVs with $P_{orb}\gtrsim P_{gap}$ and $L_{X}\sim 10^{32} \rm erg\,s^{-1}$) were found to be more concentrated than the faint CVs in these clusters, especially in the core-collapsed GCs (e.g., NGC 6397 and NGC 6752). This suggest that the bright CVs are recently formed and may suffer from dynamical encounters in the cluster cores. While for the faint CVs, they are primordial CVs or CVs that are dynamically formed a long time ago, and they may represent the highly evolved population of CVs (i.e., CVs with $P_{orb}\lesssim P_{gap}$ and $L_{X}\sim 10^{30} \rm erg\,s^{-1}$) in GCs \citep{cohn2010,lugger2017}.
Compared with faint CVs, simulations suggested that stellar encounters (such as exchange encounters) are non-negligible in the formation of bright CVs in GCs \citep{hong2017,belloni2017}, although the mass segregation effect is important in shaping the radial distribution profile of CVs in GCs \citep{belloni2018}. 

Therefore, it is likely that both primordial binaries and dynamical interactions are indispensable factors that affect the formation of weak X-ray sources in GCs. Depending on the timescale of binary evolution versus strong dynamical encounter, primordial binaries could be transformed into weak X-ray sources in GCs, either through normal stellar evolution or strong dynamical encounters. On the other hand, the weak dynamical interactions, such as the two-body relaxation, may affect the evolution tracks of primordial binaries, by driving them through sedimentation into the dense core of GCs, thus enhancing the encounter probability of binaries. We note that such a scenario is not only consistent with the observed mass segregation effect of X-ray sources in 47 Tuc, but also consistent with the recent simulation results of \citet{belloni2018}. By modeling a large sample of GCs, \citet{belloni2018} show that the majority of CVs in GCs are descendant from primordial binaries (either through primordial binary evolution channel or the dynamical formation channel of primordial binaries). More importantly, they found that the fraction of of CVs inside/outside the half-light radius is modulated by the cluster half-mass relaxation time $T_{\rm rel}$, and with longer $T_{\rm rel}$, the higher the fraction of CVs are located outside the half-light radius. Besides, bright CVs tend to be more massive and they migrate to the cluster center more faster, leading to a higher concentration than faint CVs in more dense GCs (also with smaller $T_{\rm rel}$, see Figure-9 and 10 of \citealp{belloni2018} for details).

\section{Summary}
In this work, We have presented the most sensitive and full-scale X-ray source catalogue for Globular Cluster 47 Tuc. By analyzing the radial properties of X-ray sources in the this cluster, our main findings are as follows. 

1. Our catalogue consists of 537 X-ray sources and covers a total area of $\sim 176.7$ arcmin$^{2}$ (or within a radius of $7.5\arcmin$) in 47 Tuc. 

2. The radial specific frequency of X-ray sources in 47 Tuc peaks strongly in the cluster center, rapidly decreases at intermediate radii, and finally rises again at larger radii, with two distribution dips at $R\sim 100\arcsec$ and $R\sim 170\arcsec$ for the faint ($L_{X}\lesssim 5.0\times 10^{30} {\rm\ erg\,s^{-1}}$) and bright ($L_{X}\gtrsim 5.0\times 10^{30} {\rm\ erg\,s^{-1}}$) groups of X-ray sources, respectively. These distribution shapes are similar to that of the Blue Straggler Stars (BSS), where the distribution dip is located at $R\sim 200\arcsec$ \citep{ferraro2004}.

3. By fitting the radial distribution of each heavy object (i.e., BSS, bright and faint X-ray sources) with a ``generalized King model", we estimated an average mass of $1.51\pm0.17\ M_{\odot}$, $1.44\pm0.15\ M_{\odot}$ and $1.16\pm0.06\ M_{\odot}$ for the BSS, bright and faint X-ray sources, respectively. These results are qualitatively consistent with the observed distribution dips of BSS and X-ray sources in 47 Tuc, and suggests that mass segregation plays an important role in creating these distribution features.

4. The distribution profiles of X-ray sources and BSS are consistent with the mass segregation model of binaries in GCs, which suggests that primordial binaries are a significant contributor (at least part of the contribution for the sources in cluster center and nearly a full contribution for sources located in the outskirts of 47 Tuc) to X-ray source population in GCs.

\begin{deluxetable}{lrrcrrrcrrrc}
\tabletypesize{\scriptsize}
\tablecolumns{12}
\linespread{1}
\tablewidth{0pc}
\tablenum{1}
\tablecaption{Log of {\it Chandra} Observations}
\tablehead{
\colhead{ObsID} & \colhead{Date} & \colhead{Arrays} & \colhead{Livetime} & \colhead{Ra} & \colhead{Dec} & \colhead{Roll} & \colhead{Frame}  & \colhead{$\delta x$} & \colhead{$\delta y$} & \colhead{Rotation} & \colhead{Scale}\\
\colhead{} & \colhead{} & \colhead{} & \colhead{(ks)} & \colhead{(\arcdeg)} & \colhead{(\arcdeg)} & \colhead{(\arcdeg)} & \colhead{time (s)} & \colhead{(pixel)} & \colhead{(pixel)} & \colhead{(\arcdeg)} & \colhead{Factor}\\
\colhead{(1)} & \colhead{(2)} & \colhead{(3)} & \colhead{(4)} & \colhead{(4)} & \colhead{(6)} & \colhead{(7)} & \colhead{(8)}  & \colhead{(9)} & \colhead{(10)} & \colhead{(11)} & \colhead{(12)}}
\startdata
78    & 2000-03-16 & ACIS-I & 3.87   & 5.97704 & -72.07297 & 190.4 & 0.94 & -0.078 & -0.155 &  0.028 & 1.0014 \\
953   & 2000-03-16 & ACIS-I & 31.67  & 5.97695 & -72.07304 & 190.3 & 3.24 &  0.078 & -0.230 &  0.032 & 0.9999 \\
954   & 2000-03-16 & ACIS-I & 0.85   & 5.97716 & -72.07304 & 189.9 & 0.54 &  0.269 & -0.564 & -0.088 & 0.9989 \\
955   & 2000-03-16 & ACIS-I & 31.67  & 5.97696 & -72.07294 & 189.9 & 3.24 &  0.142 & -0.281 & -0.013 & 0.9989 \\
956   & 2000-03-17 & ACIS-I & 4.69   & 5.97700 & -72.07282 & 189.5 & 0.94 &  0.098 & -0.680 &  0.035 & 0.9994 \\
2735  & 2002-09-29 & ACIS-S & 65.24  & 6.07523 & -72.08251 & 0.4   & 3.14 &  0.066 &  0.116 &  0.021 & 1.0005 \\
2736  & 2002-09-30 & ACIS-S & 65.24  & 6.07516 & -72.08274 & 359.6 & 3.14 &  0.027 &  0.161 &  0.016 & 1.0002 \\ 
2737  & 2002-10-02 & ACIS-S & 65.24  & 6.07491 & -72.08330 & 357.5 & 3.14 &  0.034 &  0.164 & -0.004 & 1.0008 \\
2738  & 2002-10-11 & ACIS-S & 68.77  & 6.07322 & -72.08552 & 349.2 & 3.14 &  0.000 &  0.000 &  0.000 & 1.0000 \\
3384  & 2002-09-30 & ACIS-S & 5.31   & 6.07515 & -72.08263 & 359.6 & 0.84 &  0.261 & -0.126 & -0.066 & 0.9985 \\
3385  & 2002-10-01 & ACIS-S & 5.31   & 6.07498 & -72.08296 & 358.8 & 0.84 & -0.028 & -0.240 &  0.038 & 1.0003 \\
3386  & 2002-10-03 & ACIS-S & 5.54   & 6.07480 & -72.08349 & 356.7 & 0.84 &  0.133 &  0.073 &  0.105 & 0.9989 \\
3387  & 2002-10-11 & ACIS-S & 5.73   & 6.07299 & -72.08544 & 349.2 & 0.84 & -0.133 &  0.028 &  0.079 & 1.0009 \\
15747 & 2014-09-09 & ACIS-S & 50.04  & 6.01653 & -72.07805 & 22.2  & 0.44 & -0.254 &  1.122 &  0.020 & 0.9987 \\
15748 & 2014-10-02 & ACIS-S & 16.24  & 6.01984 & -72.07846 & 2.2   & 0.44 & -0.955 &  1.195 &  0.216 & 1.0008 \\
16527 & 2014-09-05 & ACIS-S & 40.88  & 6.01654 & -72.07804 & 22.2  & 0.44 &  0.276 &  0.141 &  0.062 & 0.9999 \\
16528 & 2015-02-02 & ACIS-S & 40.28  & 6.01851 & -72.08395 & 236.5 & 0.44 & -0.316 &  0.300 & -0.084 & 0.9985 \\
16529 & 2014-09-21 & ACIS-S & 24.7   & 6.01989 & -72.07845 & 2.2   & 0.44 & -0.713 &  0.489 & -0.061 & 1.0005 \\
17420 & 2014-09-30 & ACIS-S & 9.13   & 6.01989 & -72.07840 & 2.2   & 0.44 & -1.168 &  0.681 &  0.007 & 1.0019 \\
\enddata
\vspace{-0.5cm}
\tablecomments{Cols. 1-3: {\it Chandra} observation ID, date and instrument layout for 47 Tuc. Cols. 4-8: Livetime, telescope optical pointing position, roll angle and frame time of each observation. Cols. 9-12: Right ascension and declination shift, rotation angle and scale factor of {\it wcs\_match} transformation matrix.}
\label{tab:obslog}
\end{deluxetable}

\begin{deluxetable}{cllccccccccc}
\tabletypesize{\scriptsize}
\tablewidth{0pt}
\tablecaption{Main {\it Chandra} Source Catalog}
\tablecolumns{12}
\linespread{1}
\tablewidth{0pc}
\tablenum{2}
\tablehead{
\colhead{XID} & \colhead{RA} & \colhead{Dec} & \colhead{Error} & \colhead{R} & \colhead{$\log P_{\rm B}$} & \colhead{$C_{net,f}$} & \colhead{$C_{net,s}$} & \colhead{$F_{p,f}$} & \colhead{$F_{p,s}$} & \colhead{$\log L_{X,f}$} & \colhead{$\log L_{X,s}$} \\
\colhead{} & \colhead{(\arcdeg)} & \colhead{(\arcdeg)} & \colhead{(\arcsec)} & \colhead{(\arcsec)} & \colhead{} & \colhead{(cts)}  & \colhead{(cts)}  & \colhead{(ph/s/cm$^{2}$)} & \colhead{(ph/s/cm$^{2}$)} & \colhead{(erg/s)} & \colhead{(erg/s)} \\
\colhead{(1)} & \colhead{(2)} & \colhead{(3)} & \colhead{(4)} &\colhead{(5)} & \colhead{(6)} & \colhead{(7)} & \colhead{(8)} & \colhead{(9)} & \colhead{(10)} & \colhead{(11)} & \colhead{(12)}
}
\startdata
1 &  5.666170 & -72.061835 & 0.5 & 402.3 & $<$-5 & $15.3^{+4.5}_{-3.9}$ & $11.5^{+3.8}_{-3.2}$ & 1.08E-6 & 5.45E-7 & 30.83 & 30.31 \\ 
2 &  5.691507 & -72.022617 & 0.3 & 424.7 & $<$-5 & $31.9^{+6.3}_{-5.5}$ & $30.7^{+6.0}_{-5.3}$ & 3.53E-6 & 2.18E-6 & 31.12 & 31.04 \\ 
3 &  5.692831 & -72.021512 & 0.2 & 425.4 & $<$-5 & $46.2^{+7.3}_{-6.7}$ & $37.1^{+6.5}_{-6.0}$ & 4.07E-6 & 2.13E-6 & 31.31 & 30.97 \\ 
4 &  5.720357 & -72.045905 & 0.2 & 359.5 & $<$-5 & $183^{+15}_{-14}   $ & $139^{+12}_{-13}$    & 4.37E-6 & 1.98E-6 & 31.28 & 30.9  \\ 
5 &  5.728573 & -72.123989 & 0.6 & 360.8 & -4.5  & $14.1^{+8.5}_{-7.9}$ & $15.4^{+6.1}_{-5.6}$ & 1.53E-7 & 9.92E-8 & 29.90 & 29.38 \\ 
6 &  5.755388 & -72.077648 & 0.4 & 297.4 & $<$-5 & $30.9^{+8.0}_{-7.4}$ & $31.4^{6.8}_{-6.1}$  & 3.78E-7 & 2.22E-7 & 30.30 & 29.78 \\ 
7 &  5.762490 & -72.120313 & 0.4 & 321.3 & $<$-5 & $26.1^{+8.3}_{-7.8}$ & $9.4^{5.0}_{-4.6}$   & 2.81E-7 & 6.02E-8 & 30.18 & 29.67 \\ 
8 &  5.765549 & -72.128034 & 0.2 & 331.4 & $<$-5 & $134^{+13}_{-13}$    & $98.8^{+10.0}_{-10.6}$ & 1.78E-6 & 7.71E-7 & 30.94 & 30.47 \\ 
\enddata
\vspace{-0.5cm}
\tablecomments{
Col.\ (1): Sequence number of the X-ray source catalog, sorted by RA.
Cols.\ (2) and (3): Right ascension and declination for epoch J2000.0.
Cols.\ (4) and (5): Estimated standard deviation of the source position error and its projected distance from cluster center.
Cols.\ (6): Logarithmic Poisson probability of a detection not being a source.
Cols.\ (7) -- (10): Net source counts and photon flux extracted in the full (0.5--8~keV) and soft (0.5-2~keV) band, respectively.
Col.\ (9) and (12): Unabsorbed source luminosity in full and soft band.
(The table is available in its entirety in machine-readable form.)}
\label{tab:mcat}
\end{deluxetable}

\begin{deluxetable}{@{}ll*{8}{c}rrr@{}}
\tabletypesize{\scriptsize} 
\tablecolumns{8}
\linespread{1}
\tablewidth{0pc}
\tablenum{3}

\tablecaption{Specific Frequence of X-ray Sources in 47 Tuc }
\tablehead{
\colhead{R} & \colhead{$N_{\rm X}$} & \colhead{$N_{\rm B}$} & \colhead{$N_{i}$} & \colhead{$N_{i}/N_{t}$} & \colhead{$L_{i}/L_{t}$} & \colhead{$R_{\rm X}$} & \colhead{$\sigma_{\rm X}$} \\
\colhead{(1)} & \colhead{(2)} & \colhead{(3)} & \colhead{(4)} & \colhead{(5)} & \colhead{(6)} & \colhead{(7)} & \colhead{(8)} }
\startdata
\multicolumn{8}{c}{Bright Group: $1.0\times 10^{-7}\lesssim F_{p}\lesssim 1.0\times 10^{-6}$}\\
\hline
0-15    & 46 & 0.20 & 45.80 & 0.213 & 0.122 & 1.75 & 0.46 \\
15-34   & 64 & 0.81 & 63.19 & 0.294 & 0.198 & 1.48 & 0.36 \\
34-51   & 33 & 1.26 & 31.74 & 0.148 & 0.143 & 1.03 & 0.31 \\
51-68   & 18 & 1.76 & 16.24 & 0.075 & 0.111 & 0.68 & 0.25 \\
68-85   & 12 & 2.27 & 9.73  & 0.045 & 0.089 & 0.51 & 0.22 \\
85-107  & 13 & 3.68 & 9.32  & 0.043 & 0.093 & 0.47 & 0.21 \\
107-127 & 12 & 4.08 & 7.92  & 0.037 & 0.069 & 0.53 & 0.25 \\
127-147 & 13 & 4.78 & 8.22  & 0.038 & 0.058 & 0.66 & 0.31 \\
147-168 & 15 & 5.77 & 9.23  & 0.043 & 0.052 & 0.83 & 0.37 \\
168-200 & 24 & 10.27& 13.73 & 0.064 & 0.065 & 0.98 & 0.38 \\
\hline
\multicolumn{8}{c}{Faint Group: $F_{p} \gtrsim 1.0 \times 10^{-6}$} \\                              
\hline
0-30    & 36 & 0.19 & 35.81 & 0.514 & 0.220 & 2.34 & 0.79 \\
30-60   & 14 & 0.58 & 13.42 & 0.193 & 0.192 & 1.00 & 0.44 \\
60-120  & 11 & 2.31 & 8.69  & 0.125 & 0.217 & 0.57 & 0.29 \\
120-220 & 8  & 7.28 & 0.72  & 0.010 & 0.184 & 0.06 & 0.01 \\
220-320 & 14 & 11.56& 2.44  & 0.035 & 0.104 & 0.34 & 0.27 \\
320-450 & 30 & 21.43& 8.57  & 0.123 & 0.083 & 1.48 & 0.76 \\
\enddata
\vspace{-0.5cm}
\tablecomments{
Col.\ (1): Annulus ranges in units of arc-second.
Cols.\ (2): Detected X-ray sources within each annulus.
Col.\ (3): Possible count of CXRB sources.
Col.\ (4): Number of background subtracted X-ray sources within each annulus, $N_{i}=N_{X}-N_{B}$.}
\label{tbl:spec_freq}
\end{deluxetable}

\begin{acknowledgements}
We thank the anonymous referee for valuable comments that helped to improve our manuscript. This work is supported by the National Key R\&D Program of China No. 2017YFA0402600, the National Science Fundation of China under grants 11525312, 11133001, 11333004 and 11303015.
\end{acknowledgements}

\label{lastpage}

\end{document}